# Carrier density modulation in graphene underneath Ni electrode

T. Moriyama, K. Nagashio*, T. Nishimura, and A. Toriumi
Department of Materials Engineering, The University of Tokyo
7-3-1 Hongo, Bunkyo, Tokyo 113-8656, Japan
*E-mail: nagashio@material.t.u-tokyo.ac.jp

We investigate the transport properties of graphene underneath metal to reveal whether the carrier density in graphene underneath source/drain electrodes in graphene field-effect transistors is fixed. The resistance of the graphene/Ni double-layered structure has shown a graphene-like back-gate bias dependence. In other words, the electrical properties of graphene are not significantly affected by its contact with Ni. This unexpected result may be ascribed to resist residuals at the metal/graphene interface, which may reduce the interaction between graphene and metals. In a back-gate device fabricated using the conventional lithography technique with an organic resist, the carrier density modulation in the graphene underneath the metal electrodes should be considered when discussing the metal/graphene contact.

## 1. Introduction

Graphene is attracting great attention as an alternative material for future high-speed field-effect transistors (FETs) because of its remarkably high carrier mobility.[1] Contact resistance, however, limits the device performance of graphene FETs.[2-5] To improve the contact resistance, a rigorous understanding of the graphene/metal interface is crucial.

Because of its monatomic, two-dimensional structure, graphene is highly sensitive to its environment. In the contact region, the properties of graphene are influenced by the contacting metal. Because of the work function difference between graphene and metals, charge transfer takes place at the interface, resulting in electrical doping in graphene. A photocurrent study has confirmed that this doped region extends hundreds of nanometers from the contact region toward the channel region because of the considerably long screening length resulting from the small density of states (*DOS*) at the Fermi level.[6] Moreover, the asymmetry in the resistance as a function of the gate voltage is reported to result from the *p-n* junction formation near the metal electrode.[7] So far, Fermi level of graphene is assumed to be fixed relative to the Dirac point (DP).[6,8]

In addition to the charge doping effect, the energy-band alteration in graphene in contact with metals has been considered. A theoretical analysis suggests that the graphene/metal interface can be classified into two groups, an adsorption group (e.g., Au, Ag, Pt) and a chemisorption group (e.g., Ni, Co, Pd).[9,10] At the chemisorption interface, it is predicted that the distance between the metal and graphene is shorter than the interlayer distance in graphite and that the *d*-orbitals of the metal are strongly hybridized with the $p_z$-orbitals of graphene,[11] resulting in the destruction of the band structure and an increase in the *DOS* in graphene. The main difference between adsorption and chemisorption metals is the degree of filling in the *d*-orbitals, which determines the stability of the antibonding states in the hybridization because a large number of electrons in the antibonding states destabilizes the hybridization.[11] Experimentally, the band structure of graphene grown on a Ni substrate has been confirmed to differ from that of the pristine graphene.[12] Therefore, the Fermi level of graphene is assumed to be fixed at the graphene/metal interface as mentioned above, especially for chemisorption metals.

However, in the conventional back-gate graphene FET device, a dependence of the contact resistivity at the Ni/graphene contact on $V_G$ has been observed,[4] suggesting that the carrier density in the graphene underneath the metal could be modulated by $V_G$, even though the graphene electrically contacts the metal.[13] This carrier density modulation was first proposed for Ti/Pd/Au contacts because the drain current at the DP increases with increasing $V_G$, even for a graphene FET device in which the entire channel region is completely covered by the topgate.[14-16] Direct experimental evidence, however, is not available at present.

In this paper, we present experimental results of a study of electrical transport in graphene just underneath Ni, a typical chemisorption metal, in graphene FET devices. The purpose of this study was to reveal whether the carrier density in graphene underneath source/drain electrodes is fixed. Moreover, the graphene/metal interaction in graphene FET devices is discussed in terms of the influence of the $SiO_2$ substrate and the organic resist residue.

## 2. Extraction of conductance modulation

First, the current flow path is considered in the device with an electrically floating metal on the graphene channel, as shown in Fig. 1. The source-drain current flows through the metal side (path (i)) because the resistivity of the metal is considerably lower than that of graphene. In this condition, it is difficult to elucidate the carrier density modulation of graphene underneath the metal. However, with decreasing metal length ($L_M$), the resistance of path (i) becomes larger than that of path (ii) due to the contribution of the graphene/metal contact resistance. This results in the preferential current flow through graphene. In this case, it is possible to detect the resistance modulation in graphene underneath the metal electrode by electrical measurements. Previously, we determined the transfer length $d_T$ ($d_T = \sqrt{\rho_C/R_{ch}}$, where $\rho_C$ is the specific contact resistivity and $R_{ch}$ is the sheet resistance of graphene underneath the metal electrode) as ~0.5-1 µm at the graphene/Ni contact region.[4] This is the effective length for the current injection at the graphene/Ni interface, as shown in Fig. 1. Therefore, the experimental condition for detecting the carrier density modulation in graphene is approximately $L_M <  2d_T$ = ~1-2 µm. This requirement can be easily met by the conventional electron-beam lithography process.

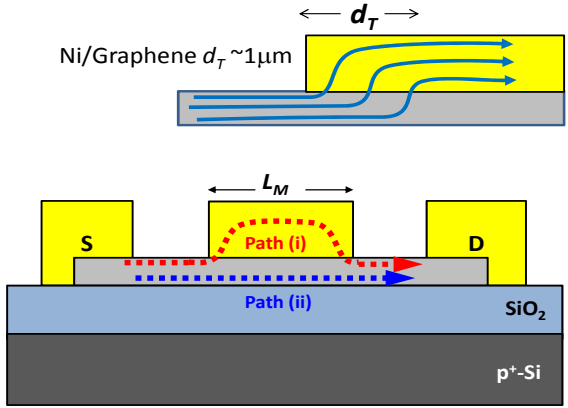

**Fig. 1** (bottom) Schematic of the device with the floating metal on the graphene channel, showing two current flow paths. (top) Schematic of the contact region, showing the transfer length ($d_T$).

## 3. Device fabrication

Monolayer graphene was transferred onto a 90-nm-thick $SiO_2$/$p^+$-Si substrate by the micromechanical cleavage of Kish graphite. The $SiO_2$/Si substrates were annealed at 1000 °C for 5 min in a 100% $O_2$ gas flow prior to the graphene transfer process. The hysteresis in the drain current - back-gate voltage curves, which may be caused by the orientation polarization of water molecules, was not observed because of the hydrophobic nature of the siloxane surface of $SiO_2$ substrates.[17] Electron-beam lithography was used to pattern electrical contacts onto graphene. The contact metal of Ni with the thickness of ~35 nm was thermally evaporated on resist-patterned graphene and subjected to lift-off with acetone. Figure 2(a) shows an optical microscope photograph of the monolayer graphene multi-terminal device structure, where the length of the Ni electrodes varies from 7.9 to 1.0 µm based on the above consideration. All Ni electrodes with various lengths were fabricated on a large "single" graphene flake. The electrical measurements were taken with a typical source-drain bias voltage of 10 mV in a vacuum. Ohmic contacts were confirmed for all electrodes.

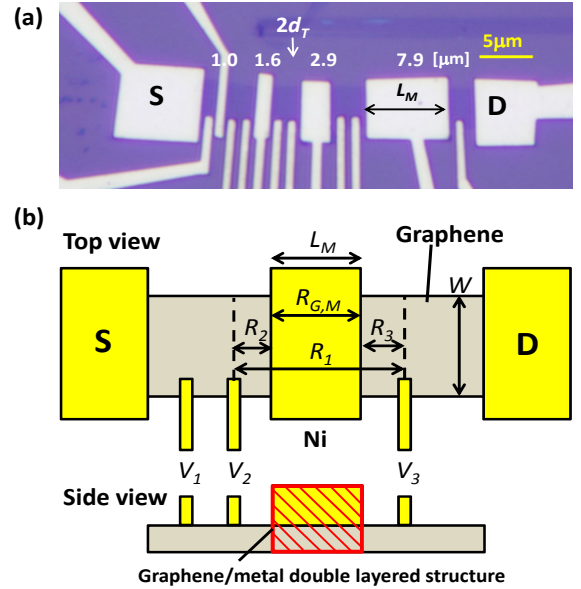

**Fig. 2** (a) Optical micrograph of the fabricated monolayer graphene FET device showing the four metal electrodes on the graphene channel. The contact metal was Ni, and ohmic contacts were confirmed for all electrodes. (b) Schematic top and side views of the device for extracting the resistivity of the graphene/metal double-layered structure.

## 4. Results and Discussion

*4.1 Carrier density modulation in graphene underneath metal*

The goal is to show the carrier density modulation of the graphene/metal double-layered structure, which is defined in Fig. 2(b). The carrier density of the metal cannot be modulated because of its large *DOS*. To analyze this device, the channel resistance ($R_1$), including both the graphene/metal double-layered structure and the intrinsic graphene regions, was initially measured by four-probe measurements using voltage probes $V_2$ and $V_3$, as shown in Fig. 2(b). Then, the resistances $R_2$ and $R_3$ were calculated from the intrinsic graphene resistivity measured by voltage probes $V_1$ and $V_2$. Finally, the resistance of the graphene/metal double-layered structure ($R_{G,M}$) could be obtained by subtracting $R_2$ and $R_3$ from $R_1$ ($R_{G,M} = R_1 - R_2 - R_3$). The mobility of this device was measured as ~4500 $cm^2V^{-1}s^{-1}$ at the carrier density of $1\times10^{12}$ $cm^{-2}$ using the four-probe measurement. The metal electrodes on the graphene channel were electrically floating in the measurement.

Figure 3(a) shows the resistivity of the intrinsic graphene without Ni ($\rho_G$) and $R_1$ for various metal lengths as a function of $V_G$. Although $\rho_G$ shows a symmetric ambipolar behavior about the DP, the shoulder shape indicated by the arrows is apparent in all $R_1$ at the hole branch. This seems to be caused by

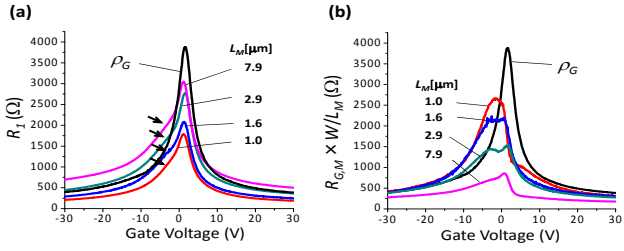

**Fig. 3** (a) Intrinsic graphene resistivity ($\rho_G$) and $R_1$ for various $L_M$ as a function of $V_G$. (b) Sheet resistivities of intrinsic graphene channel ($\rho_G$) and graphene/metal double-layers with various $L_M$ as a function of $V_G$.

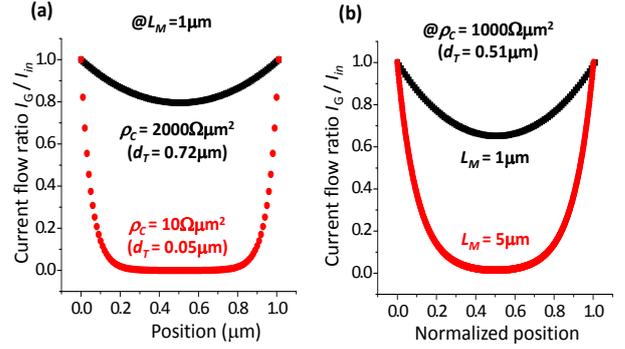

**Fig. 5** (a) Current flow ratio in graphene ($I_G/I_{in}$) calculated at $L_M$=1 μm and $V_G$= DP for various $\rho_C$. The larger $\rho_C$ prevents the current from flowing into the metal. (b) Current flow ratio in graphene ($I_G/I_{in}$) calculated at $\rho_C$=1×10³ Ωμm² and $V_G$= DP for various $L_M$. The current flow ratio in graphene increases with decreasing $L_M$.

the local shift of the DP by the charge transfer from the Ni to the graphene channel. Figure 3(b) shows $\rho_G$ and $R_{G,M}$, normalized by channel width $W$ and $L_M$, for various Ni electrodes as a function of $V_G$. As $L_M$ becomes shorter, $R_{G,M}$ clearly shows ambipolar behavior, especially for $L_M$ = 1.0 μm, which is shorter than $2d_T$. This does not mean that the degree of the carrier density modulation of graphene below the metal became more significant with decreasing $L_M$ but that the current flow ratio in the graphene increased with decreasing $L_M$. The clear ambipolar behavior is direct experimental evidence of the carrier modulation of graphene underneath the metal.

Irregular behavior was observed in $R_{G,M}$ around the DP. In particular, two peaks are evident at $V_G$ = -5 and 2 V for $L_M$ = 2.9 μm, as shown in Fig. 3(b). This irregular behavior is related with the charge transfer induced from the Ni electrode. In the $R_{G,M}$ extraction ($R_{G,M} = R_1 - R_2 - R_3$), $R_1$ includes the charge transfer region near the Ni contact, while $R_2$ and $R_3$ indeed excludes the effect of the charge transfer since $\rho_G$ without the charge transfer effect due to the external electrode structure[7] was used for the estimation. As a result, $R_{G,M}$ does include the effect of the charge transfer. Therefore, the peak observed at $V_G$ = -5 V is apparently due to the charge transfer from the Ni electrode. This small, negative shift of the peak position implies that the graphene underneath the Ni electrode is negatively doped. This cannot be explained by the simple charge transfer model based on the work function difference because of the hybridization system.

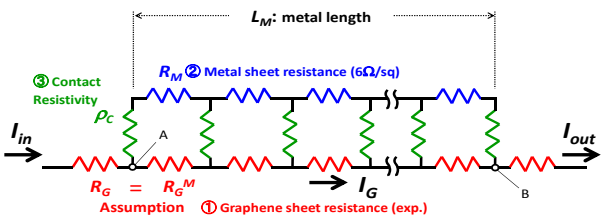

**Fig. 4** Resistor network model for the device in Fig. 2.

To examine the dependence of $R_{G,M}$ on $L_M$ more in detail, a simple resistor network model was used to calculate the current flow behavior in the graphene/metal double-layered structure. Figure 4 shows the equivalent circuit of the graphene/metal double-layered structure. The sheet resistance of the graphene underneath the metal ($R_G^M$) was assumed to be identical to the sheet resistance of intrinsic graphene ($R_G$), and the experimental value of $\rho_G$ in Fig. 3(a) was used as $R_G^M$. The DP shift is not considered in the calculation. For the sheet resistance of metal ($R_M$), we used a value of 6Ω/□, which was measured experimentally in the 35-nm-thick Ni. However, this value of $R_M$ does not affect the results at all because it is much smaller than $R_G^M$. The boundary condition is that the flow-in current ($I_{in}$) is equal to the flow-out current ($I_{out}$). In this calculation, our motivation is to show the whole picture of the current flow behavior in the graphene/metal double-layered structure, not to extract the values of physical properties precisely. Therefore, $\rho_C$ was assumed to be constant (no gate-bias dependence) just for simplicity. The $\rho_C$ and $L_M$ were used as input parameters. Then, the current for each resistance was calculated by Kirchhoff's law. The step size for $L_M$ was 0.01 μm.

Figure 5 shows the calculated current flow ratio ($I_G/I_{in}$) for various contact resistivities (a) and metal lengths (b). As expected, the current flows more preferentially through the graphene for larger $\rho_C$ and shorter $L_M$. Moreover, to reproduce Fig. 3(b), the total resistance from A to B in Fig. 4 was calculated as a function of the gate voltage for various values of $L_M$. The results are shown in Fig. 6. The values calculated for $\rho_C = 1 \times 10^3$ Ωμm² are in agreement with the experimental results as an order estimation in Fig. 3(b), supporting the hypothesis that the increase in the modulation of $R_{G,M}$ with increasing $L_M$ results from the transition of the current flow path from the metal to the graphene. In other words, the electrical property of graphene is not noticeably affected by the contact with Ni. Moreover, the multiterminal device structure in Fig. 2 was designed based on the data of $d_T$ (~0.5-1 μm) in the previous experiment.[4] The agreement between the experiment and the simple calculation also suggests that the order of $d_T$ is reasonable.

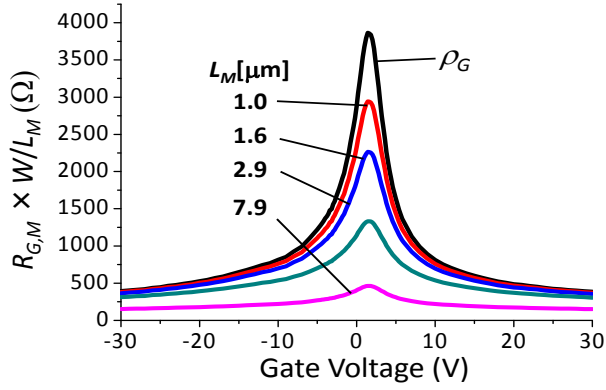

**Fig. 6** Calculated sheet resistivities for the graphene/metal double-layered structure using the experimental value of $\rho_G$, where the DP shift is not considered.

In the device in Fig. 2, the Ni electrode on graphene is electrically floating. This may be different from the grounded condition for the source electrode. Moreover, the charge transfer region near the Ni electrode is neglected in the subtraction for $R_{G,M}$ in Fig. 2; i.e., $R_2$ and $R_3$ were assumed to be the same as $\rho_G$. To exclude these uncertainties, we fabricated a different device structure using bilayer graphene with $O_2$ plasma etching, as shown in Fig. 7. The voltages in graphene underneath the metal source electrode can be directly measured using voltage probes. The voltage probe centers were positioned at 1.4, 3.8, 6.8, and 9.3 μm from the drain side of the graphene/metal double-layered structure, and the metal length on the graphene channel was 9.5 μm. The resistivities derived for each pair of voltage probes (A-C) are shown in Fig. 8(a). Because the current flows more preferentially through graphene near the drain side of the graphene/metal double-layered structure, the resistivity measured near the drain side is higher than that near the center. The smooth peak of the resistivity can be observed in Fig. 8(a), indicating that the jagged shape of $R_{G,M}$ in Fig. 3(b) may possibly originate from the subtraction method. Moreover, n-type doping in the graphene was again observed. To reproduce Fig. 8(a), calculations similar to those used to produce Fig. 6 were made. As shown in Fig. 8(b), the carrier density modulation increased as the pair of voltage probes became closer to the metal edge. When the positions of the pair of voltage probes were assumed to be 0 μm and 1 μm, the results show the pronounced carrier density modulation. This is, however, difficult to realize in an experiment. The increase in $R_{G,M}$ in Figs. 3 and 8 proves that the carrier density modulation in the graphene underneath the metal is independent of the electrode being electrically floating and grounded.

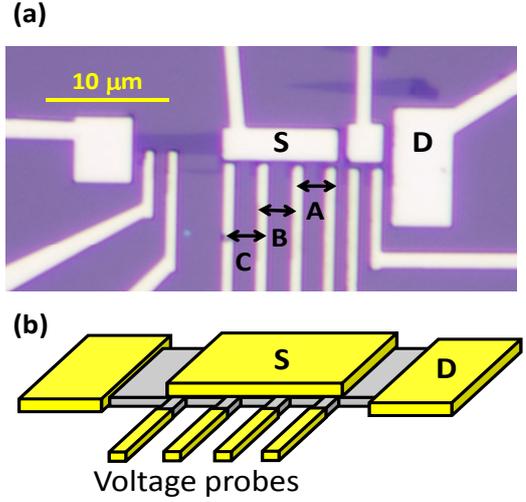

**Fig. 7** (a) Optical micrograph and (b) schematic of the bilayer-graphene FET device for direct measurements. The shape of graphene was etched by $O_2$ plasma before the metal deposition; three sets of voltage probes A~C are shown.

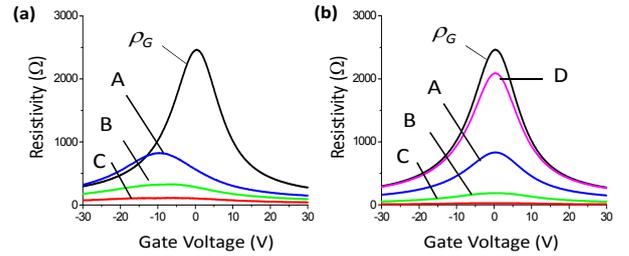

**Fig. 8** (a) Sheet resistivities of intrinsic bilayer graphene channel ($\rho_G$) and graphene/metal double-layered structure with various positions. A~C indicate measured positions in Fig. 7(b). (b) Calculated sheet resistivities for the graphene/metal double-layered structure using the experimental value of $\rho_G$, where the DP shift is not considered. D is the calculation result from the voltage difference between 0 μm and 1 μm from the metal edge.

*4.2 Effect of substrate on the Dirac point of graphene*

Previously, we reported that the different surface structures of the $SiO_2$ substrate obtained by various surface treatments result in different interactions with graphene.[17] In particular, there are two specific surface structures: the highly polarized hydrophilic silanol group (Si-OH) and the weakly polarized hydrophobic siloxane group (Si-O-Si). Because of the monatomic two-dimensional structure, the "mixed" interaction among the metal, the graphene, and the insulator, which could not be consistent with the sum of two different interactions of metal/grphene and graphene/insulator, should be considered. We have noticed negative DP shifts in Figs. 3 and 8, where the devices were fabricated on siloxane $SiO_2$ surfaces. This implies that graphene underneath the Ni electrode is negatively doped in the case of the siloxane $SiO_2$ surface. The DP shift direction for the hydrophilic $SiO_2$ surface with the same Ni electrode was the opposite; that is, hole doping occurred in the graphene (when $V_{DP}$ = ~20 V; the results are not shown). This result suggests that the doping type in graphene is affected by the "mixed" interaction among the substrate, the metal, and graphene. This may be the fundamental reason for the many inconsistent reports on the DP shift for the same metal contact, as summarized in Fig. 9.

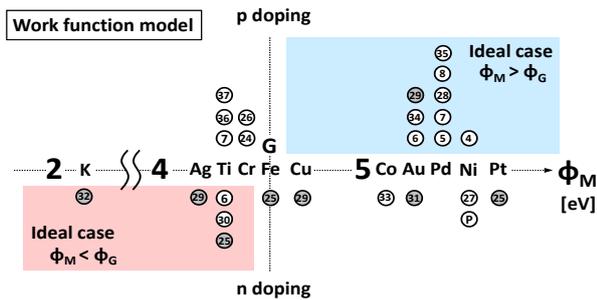

**Fig. 9** Relationship between work function for various metals and doping polarity in graphene reported in the literature. The ideal doping polarity is hatched based on the work function difference between graphene (4.5 eV) and metals. White circles indicate the doping polarity judged from the asymmetry of the current-gate voltage curve, while gray circles indicate the doping polarity judged from the DP shift when metal particles are deposited on the graphene channel. The numbers in the circles show the references.

*4.3 Effect of resist residue on carrier density modulation*

From the above results, three important points have been elucidated: 1) the carrier density in graphene can be modulated by applying a back-gate bias, even underneath the Ni electrode, 2) a Ni contact does not significantly disturb the band structure of graphene, and 3) the doping polarity in graphene also depends on the $SiO_2$ surface structure. The preservation of the band structure of graphene indicates that the orbital hybridization between graphene and Ni is not so strong, but this finding conflicts with the theoretical calculations for the metal/graphene contact region.[9,10] One possible reason for this discrepancy is the existence of resist residue at the graphene/metal interface during the device fabrication process. PMMA resist is known to be strongly attached to the graphene surface and cannot be removed completely, even after the lift-off with acetone.[18-20] This can be explained by recognizing that activated carbon, whose hydrophobic surface attracts organic materials, is composed of graphene.[21] This resist residue may alter the interaction between graphene and Ni.

First, the effect of resist residue on Ni crystallinity was investigated. A thin Ni film was directly deposited on graphene without the resist process, and the crystal orientation of the Ni on graphene was analyzed by the electron backscattering pattern (EBSP). The blue color for the normal direction of the EBSP orientation map (Fig. 10(a)) indicates that Ni(111) grew epitaxially on graphene without the annealing process, but the random color indicates that Ni was amorphous or slightly crystalline on the $SiO_2$ substrate. However, Ni was amorphous or slightly crystalline in the device structure after the conventional electron beam lithography (b). This drastic change in the Ni crystallinity suggests that the Ni atoms are affected by the honeycomb lattice of graphene in case of resist-free Ni deposition, whereas the resist residue restricts the influence of graphene in the resist-processed device case.

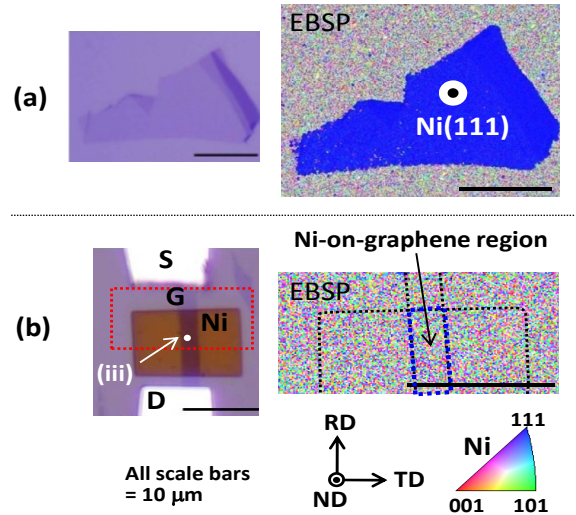

**Fig. 10** (a) Optical micrograph of graphene on the $SiO_2$ substrate. The EBSP orientation map of the normal direction (ND) is colored using the inverse pole figure triangle. Both the ND and the transverse direction (TD) show a single color, which suggests that Ni(111) grew epitaxially on the graphene. (b) Optical micrograph of the graphene FET device with a Ni electrode on the channel. The EBSP orientation map of the normal direction (ND) is colored using the inverse pole figure triangle.

Next, to verify the effect of resist residue on carrier density modulation in graphene, the Ni electrodes were deposited on graphene using the resist-free process, where a Si wafer with a 200-μm-square window was used as a metal deposition mask and the position of the Si mask was aligned with the graphene under an optical microscope. A photograph of the device with the thin (~4 nm) and thick (~30 nm) Ni electrodes is shown in Fig. 10(a). To compare fabrications with and without the resist process, another device made with the resist process was fabricated using conventional electron beam lithography, in which a PMMA resist was used. Because it was difficult to fabricate the fine electrode pattern using Si masks, a $V_G$-dependent Raman measurement was conducted to obtain the Fermi-level shift. It is well known that the Raman G-band (~1600 cm$^{-1}$) is modulated by $V_G$ because of the release from the electron-phonon coupling ascribed to the Kohn anomalies at the Γ point.[22] Figure 11(b) shows an illustration of the experimental setup. The thick Ni electrode was connected to the ground, and Raman spectra were measured through the thin Ni film as a function of $V_G$. The acquisition time for 1 spectrum was 50 s. To avoid local laser heating, the measurements were taken at a laser power of 0.8 mW under an $N_2$ gas flow. It is also noted that Ni fully covers graphene in the case where the thickness of the Ni is greater than approximately 3 nm, unlike Au.[23] Figure 12 shows the G-band position of graphene underneath the thin Ni film for samples fabricated without resist and with resist. For the resist-free graphene/Ni device, the G-band position is effectively constant, which suggests that a strong interaction between the graphene and the Ni increases the *DOS* of the graphene and prevents the Fermi level shift. In contrast, the G-band shows a clear gate bias dependence in the device fabricated with the resist process. Therefore, the resist residue at the graphene/Ni interface weakens the interaction between graphene and Ni. In the back-gate device fabricated using the conventional electron beam lithography technique, the carrier density modulation in the graphene underneath the metal electrodes should be considered when discussing the metal/graphene contact.

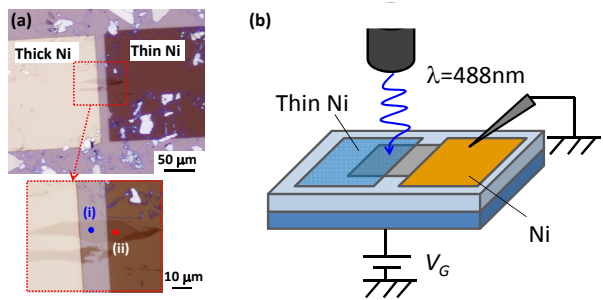

**Fig. 11** (a) Optical micrograph of the device fabricated with a resist-free process using a Si wafer mask. Thick Ni ~15 nm, thin Ni ~4 nm. (b) Schematic of the $V_G$-dependent Raman measurement system, where the thick Ni was directly contacted by the W prober tip. An objective lens (×50) with a long working distance was used.

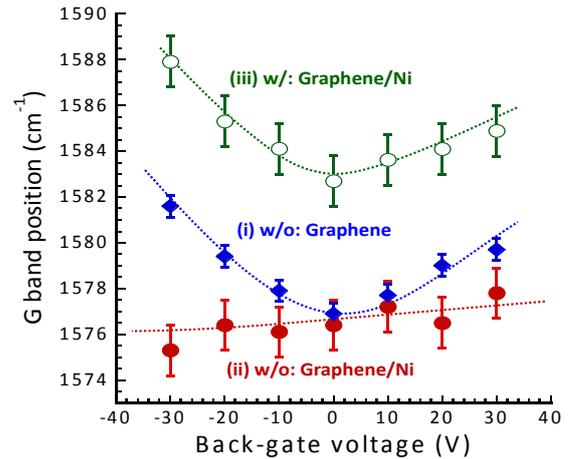

**Fig. 12** G-band position as a function of $V_G$ for different devices: (i) and (ii) for the device fabricated with a resist-free process in Fig. 11(a) and (iii) for the device fabricated with a resist process in Fig. 10(b).

## 5. Summary

We have demonstrated carrier density modulation in graphene just underneath a Ni electrode in a conventional back-gate graphene device. The ambipolar characteristics of the resistivity for the graphene/metal double-layered structure suggest that the band structure of graphene is preserved, even when in contact with the Ni. Moreover, the doping polarity in graphene underneath the Ni electrode is affected by the "mixed" interactions with the Ni electrode and the $SiO_2$ surface structures. This weak interaction between graphene and the chemisorption metal Ni can be ascribed to resist residue at the metal/graphene interface. Although the further detailed study is required, the effect of the resist residual was also observed in $V_G$-dependent Raman measurements. In a back-gate device fabricated using the conventional lithography technique, the carrier density modulation in graphene underneath the metal electrodes should be taken into account when considering the metal/graphene contact.

**Acknowledgements**

The Kish graphite used in this study was kindly provided by the Covalent Materials Co. This work was partly supported by the Japan Society for the Promotion of Science (JSPS) through its "Funding Program for World-Leading Innovative R&D on Science and Technology (FIRST Program)", a Grant-in-Aid for Scientific Research from The Ministry of Education, Culture, Sports, Science and Technology, Japan, and Semiconductor Technology Academic Research Center (STARC).